\documentclass[12pt,a4paper]{article}   
\usepackage{graphicx}                   
\usepackage{times}                      
\def\degr{$^{\circ}$}                                 
\def\arcs{$^{\prime\prime}$}                          

\title{\bf WR 140 in the Infrared}
\author{Peredur Williams\\
\vspace{1cm}\\
\normalsize Institute for Astronomy, Royal Observatory, Edinburgh, UK
\\
\\
\normalsize Published in proceedings of \\
\normalsize"Stellar Winds in Interaction", editors T. Eversberg and J.H. Knapen. \\ 
\normalsize Full proceedings volume is available on http://www.stsci.de/pdf/arrabida.pdf
}
\date{\mbox{}}
\begin{document}
\maketitle
\pagestyle{empty}
%
%
\def\bull{\vrule height .9ex width .8ex depth -.1ex}
\makeatletter
\def\ps@plain{\let\@mkboth\gobbletwo
\def\@oddhead{}\def\@oddfoot{\hfil\footnotesize\bull\quad
Workshop ``Stellar Winds in Interaction", Convento da Arr\'abida, 2010 May 29 - June 2 \quad\bull}%
\def\@evenhead{}\let\@evenfoot\@oddfoot}
\makeatother
%
%
\def\beginrefer{\section*{References}%
\begin{quotation}\mbox{}\par}
\def\refer#1\par{{\setlength{\parindent}{-\leftmargin}\indent#1\par}}
\def\endrefer{\end{quotation}}
%
%
{\noindent\small{\bf Abstract:} 
Observations in the infrared have played a significant r\^{o}le in the development of 
our understanding of WR\,140. Two sets of observations are described here: the changing 
profile of the 1.083-$\mu$m He\,{\sc i} line observed for the 2009 campaign to study 
evolution of the wind-collision region near periastron passage, and multi-wavelength 
photometry and imaging of the dust made by WR\,140 in previous periastron passages.
}
%
%
\section{Introduction}
In his contribution, Tony Moffat has described the properties of WR\,140 (= HD 193793) 
and the motivation for the 2009 campaign.
Fundamental to colliding-wind binaries is the power in their 
winds. The WC7 and O5 stars in WR\,140 have fast ($\sim$ 3000 km$^{-1}$) winds carrying 
$\sim 2 \times 10^{-5}$ and $\sim 2 \times 10^{-6} M_{\odot}$ y$^{-1}$ mass-loss respectively. 
The kinetic powers of these winds are in excess of $10^4$ and $10^3 L_{\odot}$.               
Of these, $\sim 3 \times 10^3 L_{\odot}$ is dissipated where the winds collide (the           
wind-collision region, WCR), leading to shock-heating and compression of the plasma, 
X ray emission, synchrotron radio emission, changes to the profiles 
of some emission lines and condensation of `dust' (really something like soot) in the 
shock-compressed wind. Of these effects, dust formation was certainly the most unexpected. 
The 1977 dust formation episode from WR\,140 was independently discovered (Williams et al. 
1977, 1978) at the 1.5-m Infrared Flux Collector (now the Carlos S\'anchez Telescope) 
at Iza\~{n}a on Tenerife -- only a short distance from the Mons telescope.

From 1977 to the 2001 periastron, the thermal emission from dust made by WR\,140 was
tracked with infrared photometers at a variety of wavelengths between 1 and 20 $\mu$m 
and, in the years following the 2001 dust-formation episode, with infrared cameras 
which imaged the dust clouds. These observations have been published, but I will 
summarise the most recent work (Williams et al. 2009) to show how much we can learn 
from observing in this wavelength domain. 

For the 2009 campaign, we (Watson Varricatt, Andy Adamson and the author) extended the 
observations of the 1.083-$\mu$m He\,{\sc i} line taken around the 2001 periastron 
passage (Varicatt, Williams \& Ashok 2004). This line has a P Cygni profile, with a 
flat-topped emission component like the $\lambda$5696 \AA\ C\,{\sc iii} line observed 
in the optical campaign, and develops a strong sub-peak near periastron passage. Its 
absorption component varies as we observe the stars through different parts of the wind.
The new observations were taken in 2008 with UIST on the United Kingdom Infrared Telescope 
(UKIRT) at a resolution of 200 km~s$^{-1}$ and comprise five spectra extending the coverage 
to earlier phases ($\phi$ = 0.93--0.95) and nine spectra near $\phi$ = 0.99, when the 
WR\,140 system was changing very rapidly.

\section{Geometry of the WCR from the 1.083-$\mu$m He\,{\sc i} line profile}

\begin{figure}              
\centering
\begin{minipage}{9.5cm}
\includegraphics[width=9.5cm]{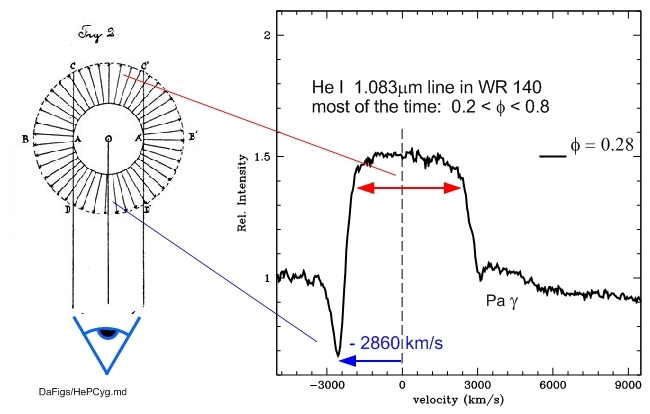}
\caption{P-Cygni profile of the He\,{\sc i} 1.083-$\mu$m line for {\em most} of the orbit. 
The weak emission feature marked Pa$\gamma$ more probably comes from the hydrogenic 
He\,{\sc ii} and C\,{\sc iv} transitions at this wavelength.}
\label{HePCyg}
\end{minipage}
\hfill
\begin{minipage}{6.5cm}
\includegraphics[width=6.5cm]{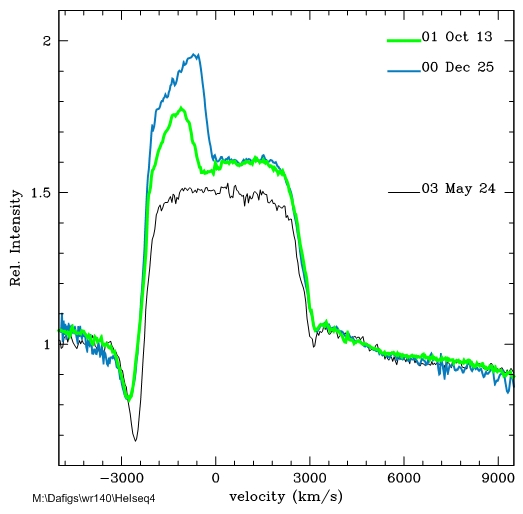}
\caption{Profiles of the 1.083-$\mu$m line showing development of the sub-peak at 
phases 0.96 and 0.986.}
\label{HeIseq4}
\end{minipage}
\end{figure}

\begin{figure}                            
\centering
\begin{minipage}{8cm}
\includegraphics[width=8cm]{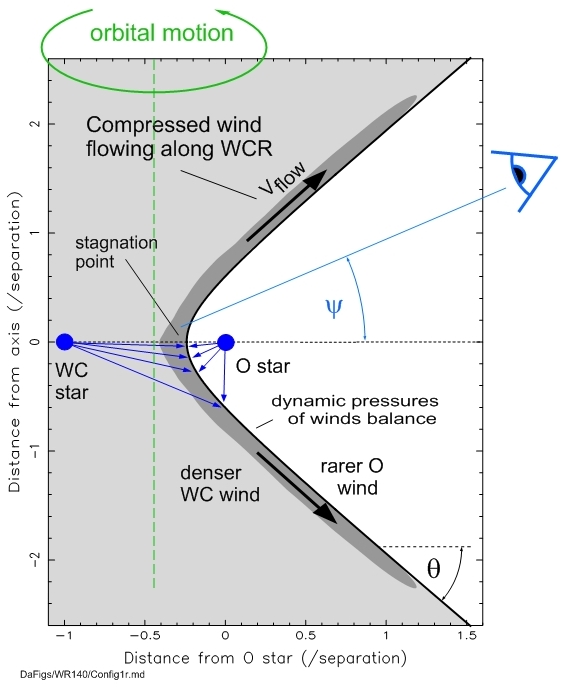}
\caption{A slice through the WCR perpendicular to the orbital plane. The stars and WCR 
rotate about the vertical axis.}
\label{Config}
\end{minipage}
\hfill
\begin{minipage}{8.5cm}
\includegraphics[width=8.5cm]{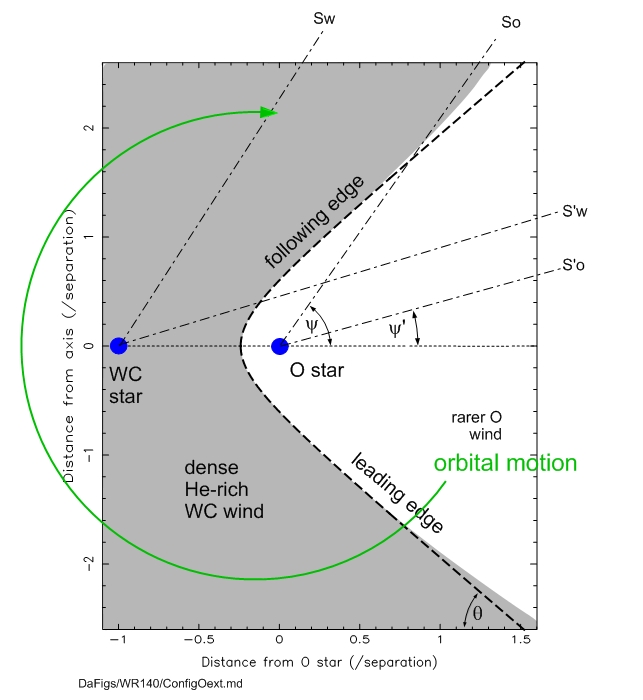}
\caption{A view of the WCR in the plane including the observer, showing how the lines of 
sight to the stars vary with orbital phase. }
\label{ConfigOext}
\end{minipage}
\end{figure}

Spectra of WR\,140 in the region of the 1.083-$\mu$m He\,{\sc i} line are shown in 
Figs \ref{HePCyg}\footnote{The sketch of an expanding stellar envelope showing where 
the absorption and emission components are formed dates from the first published 
interpretation of such a profile, in a nova spectrum in 1804, by Jakob Halm, of the 
Royal Observatory Edinburgh.} 
and \ref{HeIseq4}. The first spectra of this line in WR\,140 were taken in a programme 
to measure stellar wind terminal velocities, $v_\infty$, and not near periastron. 
The profiles in Fig.\,\ref{HeIseq4} come from the study by Varricatt et al., and show 
rapid development of the sub-peak as WR\,140 approached the 2001 periastron. 

The WCR (Fig.\,\ref{Config}) lies where the dynamic pressures of the WC7 and O5 stellar 
winds balance. 
Its shape can be calculated (e.g. Cant\'{o}, Raga \& Wilkin 1996) from the ratio of the 
momenta of the two stellar winds, 
$\eta = (\dot{M}v_\infty)_{O5}/(\dot{M}v_\infty)_{WC7}$, and is closer to the O5 star 
because its mass-loss rate and wind momentum are much lower than that of the WC star. 
At large distances from the stars, the WCR can be approximated by a cone (opening angle 
$\theta$), which is twisted to form a spiral in the orbital plane by the orbital motion. 
The pitch angle of the spiral depends on the ratio of the transverse velocity of the 
stars in their orbit relative to stellar wind velocities. In an elliptical orbit like 
that of WR\,140, this ratio varies significantly (factor $>$ 15), so the breadth of the 
spiral varies hugely round the orbit -- a bit like a snake which has swallowed a large 
animal.

The WC7 and O5 stellar winds are shocked on each side of the WCR in regions which are 
wide if the shocks are adiabatic (most of the time in the case of WR\,140) but very 
thin if the shocks are radiative (only around periastron). These shocks compress the winds, 
which flow along the WCR (only that of the WC7 star is shown in the figure). The 
compressed wind accelerates from the stagnation point to an asymptotic value, $V_{flow}$, 
on the `conical' region of the WCR, and can be calculated from the stellar winds following 
Cant\'{o} et al. It is in the compressed WC7 stellar wind, comprising mostly helium and 
carbon ions, that we believe the He\,{\sc i} and C\,{\sc iii} sub-peak emission features 
form.

What we observe in the line profile, in terms of sub-peak emission and absorption of the 
underlying stars, depends on our line of sight, which makes an angle $\psi$ with the axis 
(WC7--O5) joining the stars. This viewing angle $\psi$ varies round the orbit as 
$\cos\psi = -\sin i  \sin (f+\omega)$, where $\omega$ is determined from the RV orbit, the 
inclination, $i$, comes from other observations, e.g. an astrometric orbit, and $f$ is the 
true anomaly, the P.A.\, of the stars in their orbit. Owing to the high eccentricity of the 
orbit, the variation of $f$ with phase is very sensitive to the elements of the RV orbit. 
Work for this contribution used those of Marchenko et al. (2003), and some of the results may 
be different when they are re-calculated using the definitive orbit from the 2009 campaign.

\begin{figure}                                           
\begin{minipage}{9.5cm}
\centering
\includegraphics[width=9.5cm]{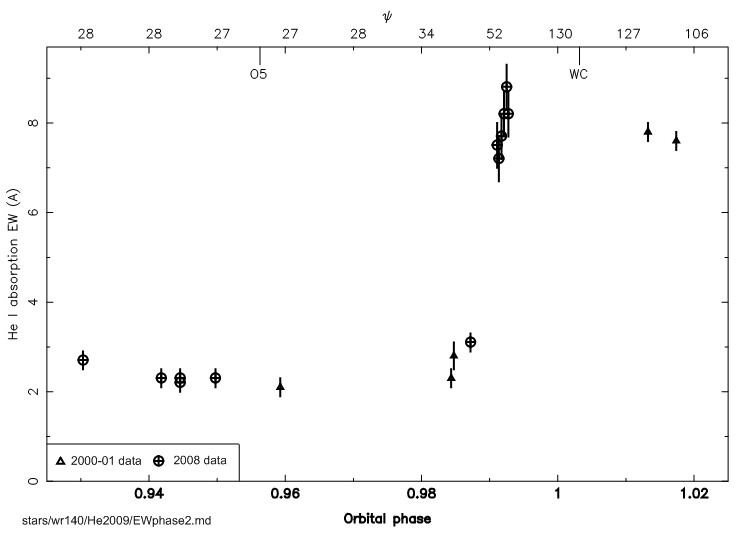}    
\caption{EW of absorption component as a function of phase near periastron. On top 
are the viewing angle, $\psi$, and phases of conjunction, `O5' and `WC', in front. }
\label{EWphase}
\end{minipage}
\hfill
\begin{minipage}{6.5cm}
\centering
\includegraphics[width=6.5cm]{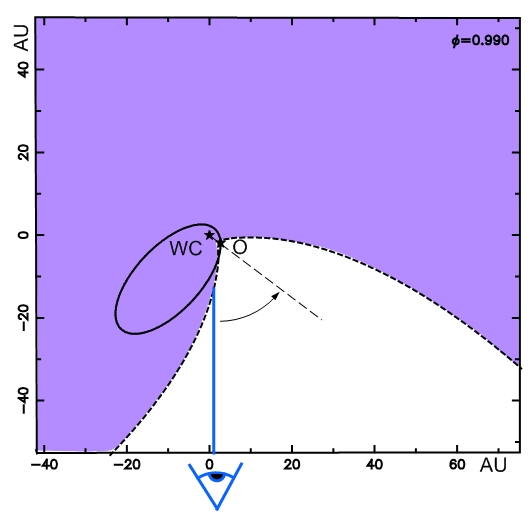}    
\caption{Sketch of the system at lower scale than Fig\,\ref{ConfigOext} for beginning of 
eclipse at $\phi$ = 0.990.}
\label{Config990}
\end{minipage}
\end{figure}

The strength of the 1.083-$\mu$m absorption varies significantly (Figs \ref{HeIseq4}  
and \ref{EWphase}). The underlying continuum comes from both the WC7 and O5 stars. 
In Fig.\,\ref{ConfigOext} we see the WCR from a different viewpoint: above the plane 
including the observer, to show how our sightlines S$^{\prime}$o and So to the O5 star 
and S$^{\prime}$w and Sw to the WC7 star change with the orbital motion. The WCR is 
asymmetric: as we move away from the stars, it lags behind the WC--O axis and cone 
because of the orbital motion, eventually to form the variable-width spiral. 
The stars and WCR move clockwise in this figure, so progressively later sightlines to 
the observer occur in the opposite sense. 
We see that the WC7 star is always viewed through some of its own 
wind, nearest the star. As this star is also the fainter component in the one-micron 
region, most of the variation observed in the 1.083-$\mu$m line absorption component 
must come from the variation in absorption to the O5 star. When the viewing angle is 
small, near conjunction ($\psi^{\prime}$ in Fig.\,\ref{ConfigOext}), the O5 star is 
seen through its own wind (sightline S$^{\prime}$o), which has one-tenth the density of 
the WC7 wind and a significantly lower helium abundance, so the absorption is at its lowest. 
This can be seen in Fig\,\ref{EWphase}, where the absorption is least near conjunction 
(O5 star in front) and barely changes when we observe it through the O5 stellar wind 
alone between the first observation and $\phi$ = 0.985, when it suddenly rises sharply. 
This must be the phase at which the following edge of the WCR crosses our sightline 
to the O5 star (Figs \ref{ConfigOext} and \ref{Config990}, i.e. when the viewing angle, 
$\psi \simeq \theta$, the cone angle, and we begin to observe the O5 star through the 
more opaque WC7 stellar wind.
Modelling the sharp increase of absorption (`eclipse') must account for the twisting of 
the WCR from the orbital motion, but we have a robust measure of the opening angle,  
$\theta$ = 50\degr\ using the Marchenko et al. RV orbit and $i$ from Dougherty et al. 
(2005). This implies a wind-momentum ratio $\eta$ = 0.10 and a comparable  
mass-loss ratio, given the similarity of the wind velocities. 


\begin{figure}                                                     
\begin{minipage}{8cm}
\centering
\includegraphics[width=7cm]{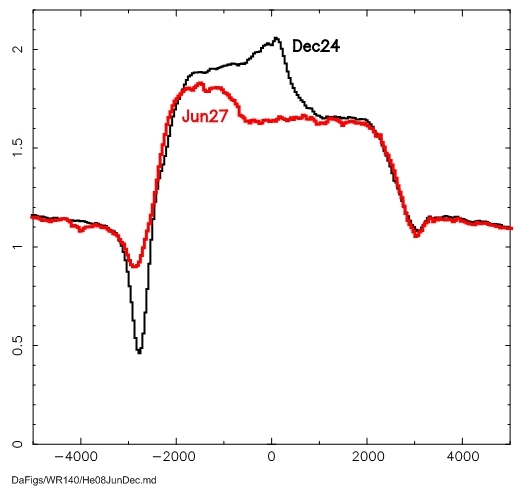}    
\caption{Spectra observed on 2008 Jun 27 and Dec 24 showing change of sub-peak profile 
as well as its movement to more positive velocity.}
\label{He08JunDec}
\end{minipage}
\hfill
\begin{minipage}{8cm}
\centering
\includegraphics[width=7cm]{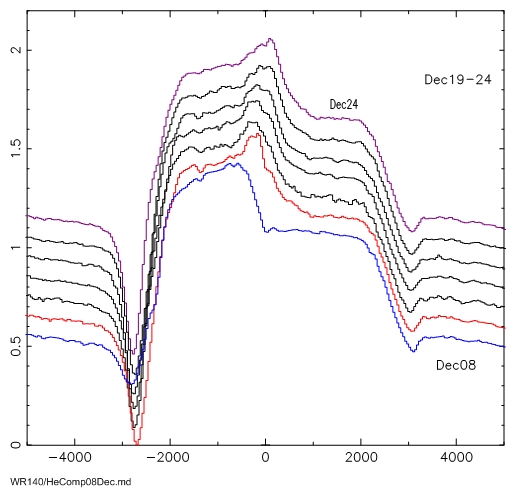}   
\caption{Montage of spectra showing evolution of sub-peak profile in 2008 December.}
\label{HeComp08Dec}
\end{minipage}
\end{figure}

\begin{figure}                                          
\begin{minipage}{9cm}
\centering
\includegraphics[width=9cm]{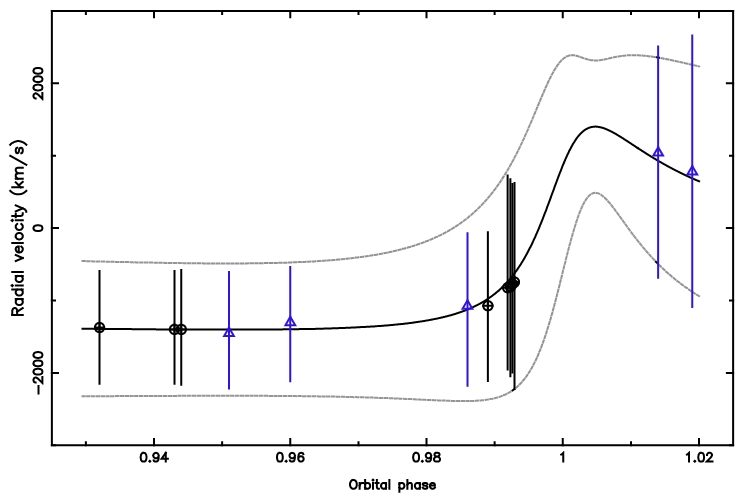}
\caption{Measured central velocities ($\oplus$, $\triangle$) and extents (vertical bars) 
of He\,{\sc i} sub-peaks compared with RVc and RVc$\pm$RVext from a L\"uhrs model.} 
\label{ThinShell}
\end{minipage}
\hfill
\begin{minipage}{7cm}
\centering
\includegraphics[width=7cm]{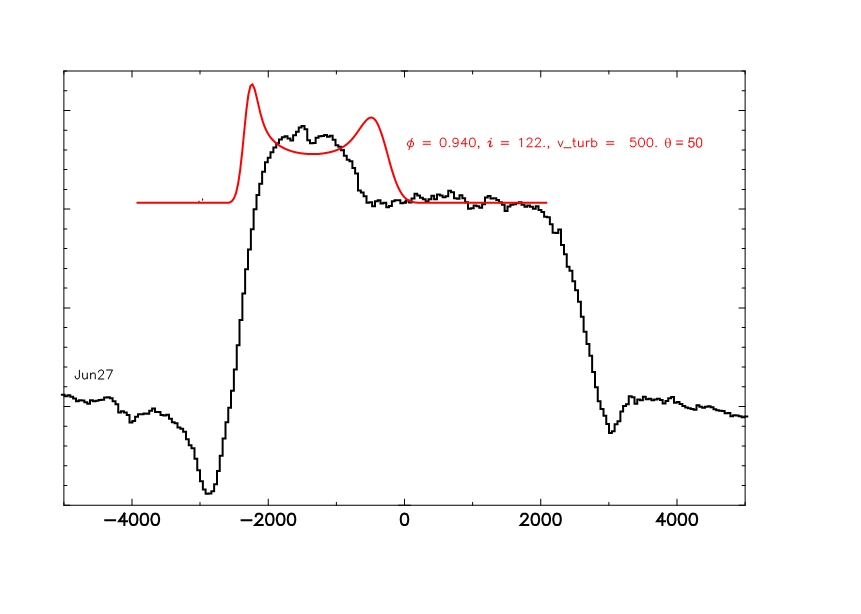}
\caption{Model profile for sub-peak profile formed in a thin shell with same parameters.
Apart from the central velocity, totally different.} 
\label{Profile}
\end{minipage}
\end{figure}

Spectra from the 2009 campaign showing the subpeak evolution are shown in Figs 
\ref{He08JunDec} and \ref{HeComp08Dec}. The first spectrum shows a blue-shifted symmetric 
profile, which barely changed as the system moved through conjunction (July and August 
spectra, not shown here). In December, the profile had shifted, broadened and become 
asymmetrical.
Modelling these changes requires knowledge of the geometry of the WCR, to be derived 
from the stellar winds, the orbit and the changes in the absorption component, the bulk 
velocity of the compressed flow (e.g. calculated following Cant\'{o} et al.) with the 
addition of an appropriate amount of turbulence, and then the line emissivity within 
the WCR, hopefully from the ever more sophisticated colliding wind models being 
calculated. 

For a first pass, we can use the simplified geometry introduced by L\"uhrs (1997), 
which considers only the conical region of the WCR, defined by $\theta$, where the 
compressed flow velocity has reached its asymptotic value, $V_{flow}$. This requires 
only two parameters, which can be constant round the orbit, and has the further 
advantage that there is no need to know the location of the line emission or its 
spatial variation in the WCR. The down-wind spiral 
shape is modelled with a single tilt angle applied to the whole structure. This 
model has been successfully applied to model the sub-peaks in several system, 
mostly having almost circular orbits. 
The observed radial velocities of the sub-peak centre, $RVc$, and extent, $\pm RVext$, 
can be related to the orbit via the viewing angle, $\psi$:
$RVc = V_{flow} \cos (\theta) \cos (\psi)$ and 
$RVext =  V_{flow} \sin (\theta) \sin (\psi)$. 

We apply this to WR\,140 using our value of $\theta$ from the absorption profile 
and $V_{flow}$ calculated following Cant\'{o} et al. and show the variation of $RVc$, 
$RVc+RVext$ and $RVc-RVext$ with phase together with $RVc$ and extents (vertical bars) 
from our observed profiles in Fig\,\ref{ThinShell}. The $RVc$ curve reproduces the 
observed central velocities reasonably well but overestimates the RV extents 
before periastron and underestimates them afterwards. But when we come to calculate 
a line profile using the same model for the WCR, the fit (Fig.\,\ref{Profile}) is 
poor. 
Tuning parameters does not improve the fit, nor is twisting of the WCR for orbital 
motion likely to help: at $\phi$ = 0.94, the transverse velocity is less than 2\% of 
the wind velocities, and even less at earlier phases, which define the spiral at the 
time of observation\footnote{Because the shape of the WCR in the orbital plane 
depends on the ratio of transverse to wind flow velocities at preceeding phases, 
the shapes of the WCR at a particular phase interval before and after periastron 
will {\em not} be the same.}. 

Models including sub-peak line formation from the curved portion of the WCR, where 
wind is still accelerating, are more promising but require knowledge of the line 
emissivity with distance from the stagnation point, which requires detailed modelling 
of conditions in the WCR. 

Returning to the observational data, it will be most interesting to compare the 
profiles of the 1.083-$\mu$m He\,{\sc i} subpeaks with those observed on the 5696\AA\ 
C\,{\sc iii} line. Do the lines have similar profiles and central velocities? Is the 
variation of their intensity with, say, stellar separation the same? This will tell 
us whether they are formed in the same region of the WCR. These profiles and line 
shifts should also be compared with those the X-ray lines observed at high resolution 
(e.g. Pollock et al. 2005 and the present campaign) to build a self-consistent model 
of the WCR.

\section{Infrared emission from dust made by WR\,140}                       

\begin{figure}                                                     
\centering
\includegraphics[width=13cm]{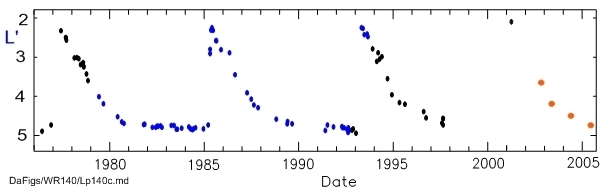}
\caption{Light curve of WR\,140 at 3.8$\mu$m; data points in orange are from IR images; 
others from photometers on UKIRT (blue) or the IRFC/TCS, SPM, TIRGO, Calgary telescopes 
or published data (black).} 
\label{Lp140}
\end{figure}

A dust grain located near a hot star with a strong stellar wind like WR\,140 experiences 
three effects: heating by the stellar radiation, accelerative force from the stellar 
radiation and collision from ions in the stellar wind. The heated grain re-radiates at 
long wavelengths with a spectrum determined from the Planck function for its temperature 
and the grain emissivity law. There have been many laboratory experiments to make analogues 
of astronomical grains and measure their optical properties, and the best fits to WR star 
dust spectra come from amorphous carbon grains. Their emissivity laws have approximately 
the same wavelength dependence in the infrared, $\kappa \propto \lambda^{-1.1}$, so the 
grain temperature can be determined observationally. The spectrum of even a relatively 
small amount ($10^{-9} M_{\odot}$) of dust can readily 
be distinguished from that of a hot stellar wind using infrared observations at a few 
wavelengths between 1 and 10 $\mu$m, and the grain temperature and, less certainly, total 
mass of dust measured. From the grain's temperature, we can determine its distance from 
the star by considering the balance between its radiative heating and re-radiation 
because the heating depends on the geometric dilution of the stellar radiation field 
at the grain's distance. The equilibrium temperature $T_{g}$ of a grain of radius $a$ 
located at a distance $r$ from a star of temperature $T_{\ast }$ and radius $R_{\ast }$ 
is related to these quantities through their Planck mean 
absorption cross-sections $\overline{Q}(a,T)$ appropriate to the grain or stellar temperature: 
\[
4\pi a^{2}\overline{Q}(a,T_{g}) T_{g}^{4}=\pi a^{2}\overline{Q}(a,T_{\ast})
\frac{4\pi R_{\ast }^{2}T_{\ast }^{4}}{4\pi r^{2}} \label{equil}
\]
For small amorphous carbon grains relevant to WR dust, $\overline{Q}(a,T) \propto aT$, 
and we get a handy relation between the equilibrium temperature of a grain and its 
distance from a star: $T_{g} \propto r^{-2/5}$. If the grain is too close to the star, 
its equilibrium temperature would be too high and it will evaporate; the corresponding 
distance is the closest at which grains can survive and is sometimes considered to be 
that at which they condense, the `nucleation radius'. 
A grain moving away from a star with the stellar wind, say, will inevitably cool and its 
infrared emission will fade. It follows that if we observe constant infrared emission 
from a WR star, as is sometimes the case, the dust being taken away in the wind must be 
being replenished by the formation of new grains at the same rate.  

In the case of WR\,140, the $L^{\prime}$ light curve (Fig\,\ref{Lp140}) shows repeated 
episodes of dust formation ($\Delta L^{\prime}$ = 2.5), followed by fading of the emission 
as the grains move away from the stars. The wavelength of this curve (3.8 $\mu$m) is 
optimal for studying WR dust formation because emission by newly formed dust peaks in 
this region while the flux from the stellar photosphere and wind have fallen off 
sufficiently by this wavelength for the dust emission to be conspicuous. 
This light curve and those in $H$ and $K$, give periods of 2905$\pm$10 d., 2905$\pm$8d. 
and 2900$\pm$10 d. respectively, consistent with the RV orbit (2899$\pm$1.3 d.). 

\begin{figure}                            
\centering
\begin{minipage}{8cm}
\includegraphics[width=8cm]{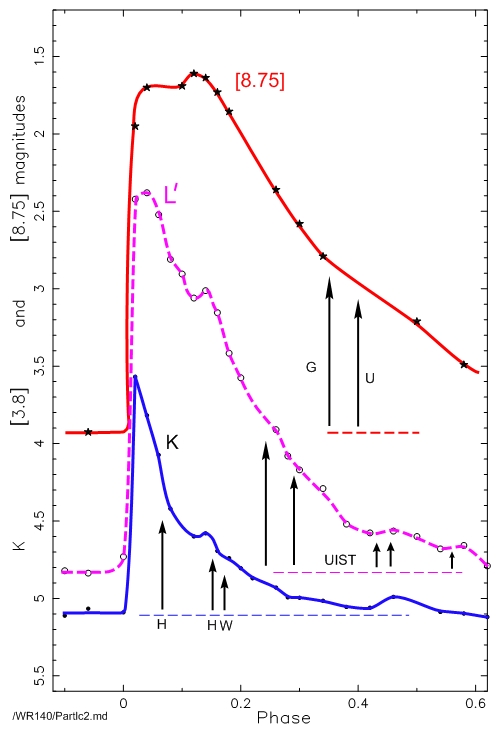}
\caption{Light curves at 8.75$\mu$m, 3.8$\mu$m ($L^{\prime}$) and 2.2$\mu$m ($K$) near 
maximum, showing also phases of imaging observations (see text).}
\label{Partlc}
\end{minipage}
\hfill
\begin{minipage}{8cm}
\includegraphics[width=8cm]{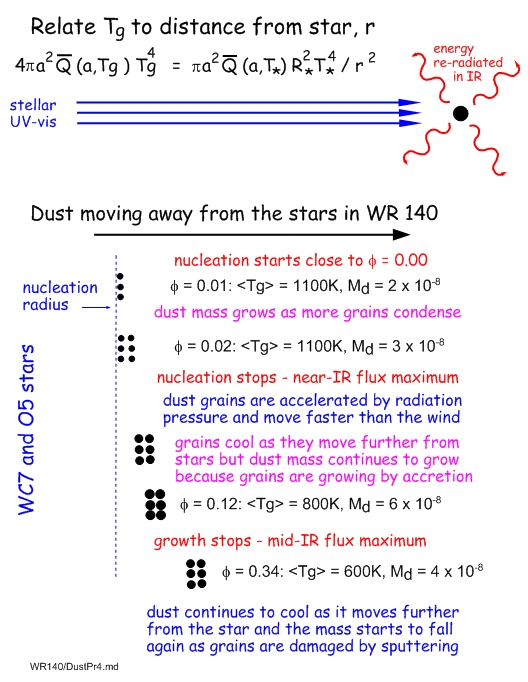}
\caption{Cartoon of processes affecting the dust grains in the wind of WR\,140}
\label{DustPr}
\end{minipage}
\end{figure}

As the fading of the IR emission is caused by cooling of the dust as it moves away 
from the stars, we expect light curves at shorter wavelengths to have steeper delines 
than those at long wavelengths, and this is largely observed (Fig.\,\ref{Partlc}), 
where we see the fading in $K$ is steeper than that in $L^{\prime}$ while that at 
8.75$\mu$m is less steep. But if this was the whole story, the light curves at all 
wavelengths would start fading at the same time, while the observations show that 
the flux maxima occur later at longer wavelengths; the 12.5-$\mu$m and 19.5-$\mu$m 
light curves resemble the 8.75-$\mu$m curve, and are even less steep.

From the multi-wavelength data, we can form spectral energy distributions (SEDs) 
for particular phases and model them to determine grain temperature and dust mass. 
The photometry tracks the integrated dust emission, so the temperature is an 
average, $\langle T_g \rangle$, of those of grains at different distances from 
the stars. 
Fits to the 1.6- to 19.5-$\mu$m and 1.6- to 12.5-$\mu$m data at phases 0.01 and 0.02 
gave $\langle T_g \rangle \simeq 1100$K and dust masses $M_d = 2\times 10^{-8}$ and 
3$\times 10^{-8} M_{\odot}$ respectively. The approximate constancy of 
$\langle T_g \rangle$, and therefore radiative-equilibrium distance from the stars, 
$r$, at these phases while the dust mass increased can be interpreted as continued 
condensation of new dust grains at a fixed 
distance from the stars, the `nucleation radius'. At $\phi = 0.00$, we have only 
$JHKL^\prime$ data and the fit to them gives $M_d \simeq 2\times 10^{-10} M_{\odot}$, 
so it is evident that dust formation had just begun at periastron. 
By $\phi = 0.03$, the $JHKL^\prime$ magnitudes were fading and $\langle T_g \rangle$ 
was falling, indicating that no new dust was condensing to replenish the dust 
carried away with the wind. From $\phi = 0.03$ to $\phi = 0.12$, the near-IR flux 
fell while the mid-IR flux rose to maximum (Fig.\,\ref{Partlc}). The model fits to 
the SEDs show that the dust mass doubled to 6$\times 10^{-8} M_{\odot}$ while 
$\langle T_g \rangle$ fell to 800~K at $\phi$ = 0.12. This shows that the increase 
of dust mass was caused by the growth of the recently formed grains at their 
radiative equilibrium temperature rather than by the condensation of fresh 
grains at $T_g \simeq 1100$K. 
This is summarised in a cartoon in Fig.\,\ref{DustPr}. 
The grains can grow by implantation of carbon ions they collide with in the wind 
and this process is greatly helped if the grains are forced to move through the 
wind by radiation pressure -- which they certainly experience. Evidence for larger 
grains made in WR\,140 comes from the eclipses observed in the optical light 
curves between phases 0.020 and 0.055 by Marchenko et al.

The $K$ and $L^\prime$ light curves show an inflexion at $\phi = 0.14$, suggesting a 
short-lived increase in dust emission. The $H$ photometry does not show any interuption 
in its fading at this phase, suggesting that the brightening at $K$ and $L^\prime$ 
does not come from the condensation of new, hot dust but from a temporary increase in 
the growth rate of the grains. 

After $\phi = 0.14$, the fading continues at all wavelengths and the dust 
cools as expected. The total mass, however, falls from its maximum of 
6.5$\times 10^{-8} M_{\odot}$ to less than 3$\times 10^{-8} M_{\odot}$. 
It suggests that, as grains move through the wind, the rate of destruction 
by thermal sputtering eventually overtakes that of growth by implantation 
of carbon ions and grains are destroyed.

\begin{figure}                                            
\centering
\includegraphics[width=12cm]{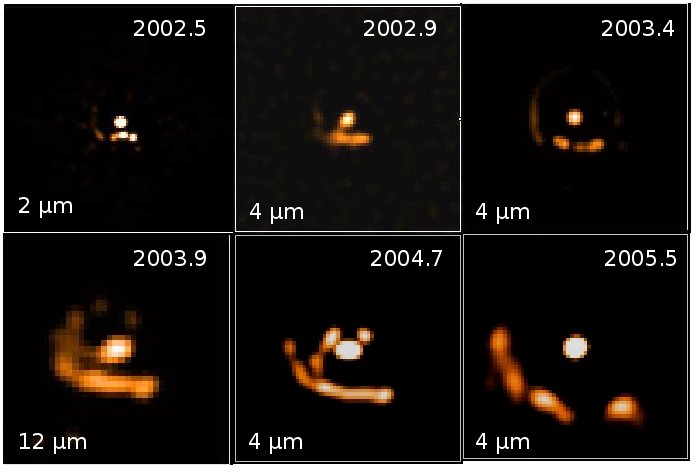}
\caption{Infrared images of WR\,140 all on the same scale, 4\arcs square, with North 
at the top, East to the left, with dates and wavelengths of observations.} 
\label{Montage}
\end{figure}

The light curves give invaluable information on the integrated properties of the 
dust, including its radial expansion, and are nicely complemented by imaging 
observations, which map the dust on the sky and allow us to relate it to the orbit. 
The dust emission by WR\,140 was first imaged in 2001 (phases 0.04 and 0.06) by 
Monnier, Tuthill \& Danchi (2002) with the Keck telescope. 
More images at wavelengths between 2.2$\mu$ and 12.5$\mu$m were observed in a
multi-site campaign by Sergey Marchenko, Tony Marston, Tony Moffat, Watson 
Varricatt and the author using the Hale 5-m telescope, William Herschel Telescope 
(WHT), UKIRT and Gemini North with infrared cameras and, in some cases, adaptive 
optical systems. As the dust cooled, it was necessary to observe it at longer 
wavelengths. Details are given by Williams et al. (2009). The phases of the 
observations are marked on the light curves closest in wavelength to the images in 
Fig.\,\ref{Partlc} with `H' (Hale), `W' (WHT), `U or UIST' (UKIRT) and `G' (Gemini).
A montage of images, all on the same scale, is presented in Fig.\,\ref{Montage}.
We see dust features to the S and E of the stars (which are not resolved), 
expending away from them. The features look like clumps but are better described 
as concentrations in extended emisssion with brightness enhanced by the maximum 
entropy reconstruction processing of the images. (The two compact features close 
to the stars in the 2004.7 image are instrumental artefacts.) 

\begin{figure}                            
\centering
\begin{minipage}{7cm}
\includegraphics[width=7cm]{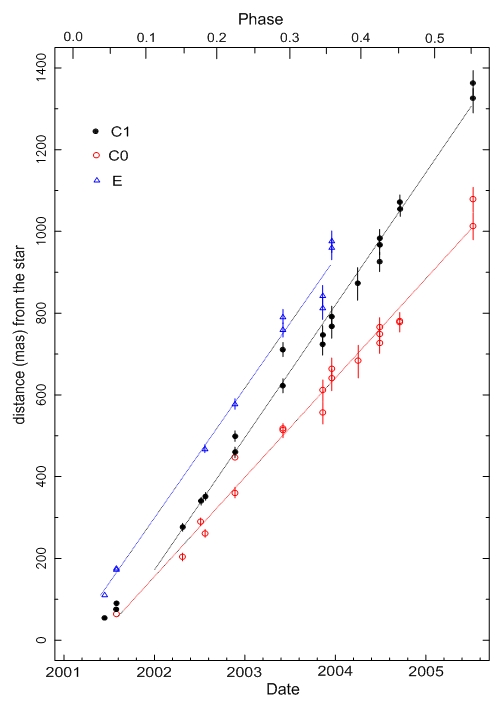}
\caption{Radial distances (mas) of dust features (see text) against date and phase.}
\label{DustMotion}
\end{minipage}
\hfill
\begin{minipage}{9cm}
\includegraphics[width=9cm]{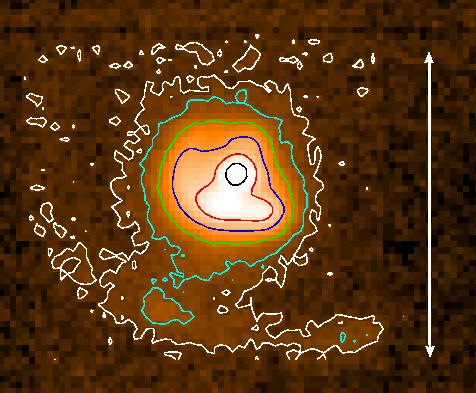}
\caption{Combined 2003 November+December 12.5-$\mu$m image (5\arcs scale marker) 
at a more sensitive intensity scale than Fig.\,\ref{Montage} to show the faint 
emission from the 1993 dust-formation epsiode to the south-east and south-west 
of the 2001 dust feature.}
\label{OldDust}
\end{minipage}
\end{figure}

The positions of persistent features, `C' at the west end of the `bar' to the south 
of the stars, `C0' at the east end of the `bar' and `E' in the feature to the east,
have been measured. The position angles of each are constant, showing radial 
motion, and the distances are plotted against date and phase in Fig.\,\ref{DustMotion}.
Within the errors, the proper motions are constant, suggesting that the dust is moving 
into a `clean' environment, presumably cleared by the O5 and WC stellar winds. Some
of the long-wavelength images (Fig.\,\ref{OldDust}) show dust features corresponding 
to `C' and `C0' but further away -- at distances which fit the extrapolated proper 
motions derived from the 2001 dust! 

The constancy of the proper motions seems surprising given that the grains must 
experience radiation pressure: a 100\AA\ located 100 AU from the O5 and WC stars 
will gain 100 km s$^{-1}$ in velocity within one week. But as a grain accelerate 
to move faster than the wind, it experience supersonic drag, which increases 
with the square of the velocity relative to the wind until the radiation pressure 
is balanced by the drag and the grain has a constant, `drift', velocity relative 
to the wind. The expansion velocity of the dust is then the sum of $V_{flow}$, 
inherited from the compresssed wind, and the drift velocity. Projection of the 
proper motions back to the stars suggests that feature `E' was formed about 
136~d. (0.05~P) before `C' and `C0'.

The images allow us to separate photometry of the dust features from that of the 
stars and, when we have images at two wavelengths (e.g. 3.6$\mu$ and 3.99$\mu$m on 
UKIRT or 7.9$\mu$m and 12.5$\mu$m on Gemini), we can determine infrared colours.
The results show that the dust features are `redder' than the stars in 
([3.6]--[3.99]), as expected from heated dust and stellar wind emission, but 
`bluer' in ([7.9]--[12.5]). The latter may be surprising but is correct: at 
the time of the observations, the dust emission peaked at a wavelength shorter 
than 7.9$\mu$m, and the Rayleigh-Jeans tail of the Planckian dust spectrum is 
steeper than the spectrum of the free-free emission from the stellar wind.

Relation of the dust maps to the projected orbit is difficult. From combination of 
their radio images with the RV orbit, Dougherty et al. derived $i$ and $\Omega$, 
orienting the orbit on the sky. They noted that the O5 star was NW of the WC7 star 
at the time of periastron, and commented on the paucity of dust in that direction 
given that the base of the WCR (i.e. the open end), where dust was expected to 
form, would have been pointing in that direction. Owing to the high orbital 
eccentricity, however, the position angle of the WCR changes very rapidly around 
periastron passage (e.g. the P.A. of the O5 star relative to the WC star moves 
through three-quarters of its orbit in only 0.04P. These ranges are very sensitive 
to the orbital eccentricity itself, because it is so high, and account for the 
spreading of the dust around much of the orbit despite the short duration of dust 
formation. Also, the density along the dust plume varies sharply with P.A. Even if 
dust forms at a constant rate, the density on the sky of dust condensed from the 
compressed wind originating at periastron is more than ten times lower than that 
originating at phase 0.02, where the WCR sweeps round more slowly, and the dust 
is spread less thinly. 

Models for the dust emission have been constructed using the orbit, timing of dust 
formation from the IR light curves, $V_{flow}$ and the shape of WCR from the winds, 
populating the down-stream WCR with dust. This is projected on the sky to get images 
for comparison with the observed images. The model images are very sensitive to the 
orbital elements, especially eccentricity, and the fits are improved by dropping 
the assumption that the wind densities in the WCRs are uniform around their axes 
and having higher densities on the `trailing’ edge in the plane, which is consistent 
with with some models of WCRs. This also accounts for the lower proper motion of 
one of the dust features (`C0', Fig.\,\ref{DustMotion}).

Successful modelling of the dust formation by WR\,140 is the biggest test of the 
colliding-wind dust-formation paradigm because we know so much about this system; 
if we cannot model it, we cannot claim to understand dust formation by the CWBs. 
And, in any fitting process, we should remember the words of John von Neumann 
(as quoted by Enrico Fermi) ``with four parameters I can fit an elephant and 
with five I can make him wiggle his trunk.''

%
%
\section*{Acknowledgements}
The author is grateful to the Institute for Astronomy for hospitality 
and continued access to facilities of the Royal Observatory, Edinburgh. 
%
%
\footnotesize
\beginrefer

\refer Cant\'o J., Raga A. C., Wilkin F. P., 1996, ApJ, 469, 729

\refer Dougherty S. M., Beasley A. J., Claussen M. J., Zauderer A., Bolingbroke N. J., 
              2005, ApJ 623, 447

\refer L\"uhrs S., 1997, PASP, 109, 504

\refer Marchenko S. V., et al., 2003, ApJ, 596, 1295


\refer Monnier J. D., Tuthill P. G., Danchi W. C., 2002, ApJ, 567, L137

\refer Pollock A. M. T., Corcoran M. F., Steven I. R., Williams P. M.,2005, ApJ 629, 482  

\refer Varricatt W. P., Williams P. M., Ashok N. M. 2004, MNRAS, 351, 1307   

\refer Williams P. M., Stewart J. M., Beattie D. H., Lee T. J., 1977, IAU Circ. 3107, 2 

\refer Williams P. M., Beattie D. H., Lee T. J., Stewart J. M., Antonopolou E., 1978, MNRAS 185, 467 


\refer Williams P. M., Marchenko S. V., Marston A. P., et al. 2009, MNRAS 395, 1749

\endrefer           
\end{document}